\documentclass[notitlepage,openbib,12pt]{article}
\usepackage{amsfonts}
\usepackage{amsmath}
\usepackage{amssymb}
\usepackage{graphicx}
\usepackage{color}
\usepackage[colorlinks,citecolor=blue]{hyperref}
\usepackage{natbib}
\usepackage{mathpazo}
\usepackage{amssymb}

\newtheorem{theo}{Theorem}

\newtheorem{define}{Definition}

\newtheorem{lemma}{Lemma}

\makeindex
\input epsf

\begin{document}

\title{\textbf{B{\"o}rgers's Open Question Resolved}}
\author{Siyang Xiong\thanks{
Department of Economics, University of California, Riverside, United States,
siyang.xiong@ucr.edu}}
\maketitle

\begin{abstract}
Focusing on stochastic finite-action mechanisms, we study implementation in
undominated strategies and iteratively undominated strategies. We establish
both possibility and impossibility results that resolve the open question in 
\cite{tb95}. Contrary to the conventional understanding that positive
results on Nash implementation need separability, quasilinearity, or
infinite action sets, we provide---to our knowledge---the first positive
result beyond those demanding assumptions.
\end{abstract}

\newpage

\baselineskip= 20pt

\section{Introduction}

\label{sec:intro}

\cite{tb95} proved a remarkable impossibility result three decades ago:
under mild domain restrictions, implementation in undominated
strategies---and likewise implementation via iterative deletion of dominated
strategies---by deterministic (finite-action) mechanisms necessarily yields
dictatorial social choice functions. As \cite{tb95} emphasized, the proof
leans crucially on a discrete feature of deterministic mechanisms. Once
lotteries are allowed, that discreteness vanishes. This led \cite{tb95} to
raise a long-standing open question: does the impossibility still hold when
mechanisms may be stochastic?

This question is important for several reasons. First, both undominated
strategies and their iterative counterpart are fundamental solution
concepts; implementation under these notions is therefore of independent
interest. While implementation via iterative procedures has been studied,
prior work typically alters either the implementation notion (e.g., \cite%
{am1992, dahm1992}, \cite{bm2009}) or the underlying solution concept (e.g., 
\cite{am1994}, \cite{bmt}) or primitive setups (e.g., \cite{cksx}), so \cite%
{tb95}'s specific question has remained unanswered. Second, most classic
impossibility results are proved on unrestricted domains (e.g., Arrow's
theorem (\cite{ka1951}); Gibbard--Satterthwaite theorem (\cite{Gibbard1973,
gibbard78}, \cite{ms1975})). By contrast, \cite{tb95} obtains impossibility
on \emph{any} restricted domain satisfying a mild condition, suggesting that
undominated-based implementation may be robustly impossible. Third, for
finite-action mechanisms, iterative deletion of strictly dominated
strategies coincides with iterative deletion of never-best strategies---the
notion commonly called rationalizability. By contrast, with infinite action
sets the two procedures can diverge (see \cite{bl1994}). When \cite{bmt}
introduce rationalizable implementation with stochastic infinite-action
mechanisms, they adopt the \textquotedblleft never-best\textquotedblright\
version. One could instead define rationalizable implementation via
iterative deletion of dominated strategies. If \cite{tb95}'s impossibility
result extends to stochastic mechanisms, then the non-trivial
rationalizable-implementation results in \cite{bmt} would hinge on that
choice of rationalizability rather than on substantive possibilities per se.
Thus, clarifying the logical relationships between these definitions---and
pinpointing where they part ways---is of first-order importance; resolving 
\cite{tb95}'s open question is a natural first step.

To date, the only work that directly addresses \cite{tb95}'s question is 
\cite{ks11}. They propose a novel necessary condition for implementation in
iteratively undominated strategies. However, this new necessary condition
cannot help us answer \cite{tb95}'s open question. As they write:

\begin{quotation}
\cite[page 2594]{ks11}: \textit{However, the good news stop there. As
already pointed out by B\"{o}rgers [6], the use of stochastic mechanisms
seems to create real difficulties to the rest of the argument. We have not
been able to prove the extension of the result, but we have not been able to
find a counterexample either. In this regard, the connections with Majumdar
and Sen [11] and their impossibility result for ordinally Bayesian incentive
compatibility seem quite relevant. We leave this intriguing issue as an
important open question.}
\end{quotation}

This paper focuses on stochastic finite-action mechanisms and provides both
negative and positive answers to \cite{tb95}'s question. First, for
undominated strategies ($UD$), implementation is necessarily dictatorial
(Theorem \ref{thm:UD:finite}). Second, for iteratively undominated
strategies ($UD^{\infty }$), if there are only two social outcomes,
implementation remains dictatorial (Theorem \ref{thm:UD2:2-outcomes}).
Third, once there are three or more outcomes, non-dictatorial social choice
functions can be implemented under $UD^{\infty }$ by a finite-action
stochastic mechanism (Lemma \ref{lem:Z:3:UD-finite} and Theorem \ref%
{thm:UD2:3-outcome}). We also study infinite-action mechanisms: they enlarge
the scope of $UD$ implementation (Theorem \ref{thm:UD:infinite}) but do not
overturn the two-outcome $UD^{\infty }$ impossibility.

\bigskip

Perhaps the most surprising result is the positive $UD^{\infty }$
construction with three or more social outcomes (Theorem \ref%
{thm:UD2:3-outcome}). A prominent critique of implementation theory targets
\textquotedblleft integer/modulo games\textquotedblright\ and related
pathologies (absence of best replies, problematic mixed equilibria),
motivating \cite{Jackson1992Bounded}'s program of implementation with
finite-action mechanisms. To date, positive results in that program
typically require strong assumptions (e.g., separable environments, \cite%
{mg2011}, \cite{cmlr2}, \cite{cksx, cksx2022}). Theorem \ref%
{thm:UD2:3-outcome} is, to our knowledge, the first finite-action positive
implementation result that dispenses with such strong assumptions.

Furthermore, one implication \cite{ks11} draw from their result is:

\begin{quotation}
\cite[page 2583]{ks11}: \textit{We conclude that either the
quasilinearity-like assumptions of available sufficiency results cannot be
completely dispensed with or some mechanisms that do not satisfy the best
element property must be employed. }
\end{quotation}

In contrast, Theorem \ref{thm:UD2:3-outcome} shows that we can dispense with
quasilinearity-like assumptions and implement non-dictatorial SCFs by
finite-action mechanisms that satisfy the best element property in \cite%
{ks11}. Stochastic mechanisms thus offer more design leverage than
previously thought.

The remainder of the paper proceeds as follows: we define the model in
Section \ref{sec:model}; we review \cite{tb95} in Section \ref{sec:Borgers};
we study undominated strategies and iteratively undominated strategies in
Sections \ref{sec:impossibility:UD} and \ref{sec:impossibility:UD-infinite},
respectively; we consider infinite-action mechanisms in Section \ref%
{sec:infinite-action}; we conclude in Section \ref{sec:conclude}.

\section{\ Model}

\label{sec:model}

We consider a finite set of agents $\mathcal{I}$ with $\left\vert \mathcal{I}%
\right\vert \geq 2$, and a finite set of social outcomes $Z$ with $%
\left\vert Z\right\vert \geq 2$. An ordinal state, denoted by $\theta \equiv
\left( \succeq _{i}^{\theta }\right) _{i\in \mathcal{I}}$, is a profile of
agents' complete and transitive preferences over $Z$. Let $\Theta ^{\ast }$
denote the set of all possible ordinal states.

For each $\theta \in \Theta ^{\ast }$, we say $u^{\theta }\equiv \left(
u_{i}^{\theta }:Z\longrightarrow 
\mathbb{R}
\right) _{i\in \mathcal{I}}$ \ is a cardinal representation of $\theta
\equiv \left( \succeq _{i}^{\theta }\right) _{i\in \mathcal{I}}$ if and only
if%
\begin{equation*}
z\succeq _{i}^{\theta }z^{\prime }\Longleftrightarrow u_{i}^{\theta }\left(
z\right) \geq u_{i}^{\theta }\left( z^{\prime }\right) \text{, }\forall
\left( z,z^{\prime },i\right) \in Z\times Z\times \mathcal{I}\text{.}
\end{equation*}%
$U_{i}^{\theta }$ is the expected utility function associated with $%
u_{i}^{\theta }$: for all $(y,i,\theta )\in \Delta (Z)\times \mathcal{I}%
\times \Theta $, 
\begin{equation*}
U_{i}^{\theta }(y)\equiv \sum_{z\in Z}y_{z}\times u_{i}^{\theta }(z),
\end{equation*}%
where $y_{z}$ denotes the probability assigned to $z$ under lottery $y$.
With slight abuse of notation, we regard a deterministic outcome as a
degenerate lottery, i.e., $Z\subseteq \Delta (Z)$. We write $-i$ for $%
\mathcal{I}\diagdown \{i\}$.

Each $\left( u_{i}^{\theta }:Z\longrightarrow 
\mathbb{R}
\right) _{i\in \mathcal{I}}$ is called a cardinal state. Define%
\begin{equation*}
\Omega ^{\left[ \theta ,\text{ }%
\mathbb{R}
\right] }\equiv \left\{ u^{\theta }\equiv \left( u_{i}^{\theta
}:Z\longrightarrow 
\mathbb{R}
\right) _{i\in \mathcal{I}}:%
\begin{tabular}{l}
$z\succeq _{i}^{\theta }z^{\prime }\Longleftrightarrow u_{i}^{\theta }\left(
z\right) \geq u_{i}^{\theta }\left( z^{\prime }\right) \text{,}$ \\ 
$\forall \left( z,z^{\prime },i\right) \in Z\times Z\times \mathcal{I}\text{.%
}$%
\end{tabular}%
\right\} \subseteq \left( \left( 
\mathbb{R}
\right) ^{Z}\right) ^{\mathcal{I}}\text{, }\forall \theta \in \Theta ^{\ast }%
\text{.}
\end{equation*}%
That is, $\cup _{\theta \in \Theta ^{\ast }}\Omega ^{\left[ \theta ,\text{ }%
\mathbb{R}
\right] }$ is the set of all possible cardinal states.

A stochastic mechanism (or simply, a mechanism) is a tuple $M=\langle
S=\times _{i\in \mathcal{I}}S_{i},\ g:S\longrightarrow \Delta (Z)\rangle $,
where $S_{i}$ is a finite strategy set for agent $i$.\footnote{%
We mainly focus on finite-action mechanisms, and will consider
infinite-action mechanisms only in Section \ref{sec:infinite-action}.} Given
any $\widetilde{S}\equiv \times _{j\in \mathcal{I}}\widetilde{S}%
_{j}\subseteq S$, we say $s_{i}\in S_{i}$ is strictly dominated on $%
\widetilde{S}$ in $M$ at a cardinal state $u^{\theta }\equiv \left(
u_{j}^{\theta }\right) _{j\in \mathcal{I}}$ if and only if\footnote{%
Here, we restrict attention to pure strategies dominating other strategies.
One could alternatively allow domination by mixed strategies. For
simplicity, we adopt the former; however, all of our results continue to
hold under the latter.}%
\begin{equation*}
\exists s_{i}^{\prime }\in S_{i}\text{, }U_{i}^{\theta }\left[ g\left(
s_{i}^{\prime },s_{-i}\right) \right] >U_{i}^{\theta }\left[ g\left(
s_{i},s_{-i}\right) \right] \text{, }\forall s_{-i}\in \widetilde{S}_{-i}%
\text{.}
\end{equation*}%
Define%
\begin{equation*}
UD_{i}^{1}\left( M,\text{ }u^{\theta }\right) \equiv \left\{ s_{i}\in
S_{i}:s_{i}\text{ is not strictly dominated on }S\text{ in }M\text{ at }%
u^{\theta }\right\} \text{, }\forall i\in \mathcal{I}\text{,}
\end{equation*}
and recursively, for any positive integer $k$, define%
\begin{equation*}
UD_{i}^{k+1}\left( M,\text{ }u^{\theta }\right) \equiv \left\{ s_{i}\in
UD_{i}^{k}\left( M,\text{ }u^{\theta }\right) :%
\begin{tabular}{l}
$s_{i}\text{ is not strictly dominated}$ \\ 
$\text{on }\times _{j\in \mathcal{I}}UD_{j}^{k}\left( M,\text{ }u^{\theta
}\right) \text{ in }M\text{ at }u^{\theta }$%
\end{tabular}%
\text{ }\right\} \text{, }\forall i\in \mathcal{I}\text{.}
\end{equation*}%
Define%
\begin{equation*}
UD\left( M,\text{ }u^{\theta }\right) \equiv \times _{i\in \mathcal{I}%
}UD_{i}^{1}\left( M,\text{ }u^{\theta }\right) \text{ and }UD^{\infty
}\left( M,\text{ }u^{\theta }\right) \equiv \times _{i\in \mathcal{I}}\left[
\cap _{k=1}^{\infty }UD_{i}^{k}\left( M,\text{ }u^{\theta }\right) \right] 
\text{.}
\end{equation*}%
That is, $UD\left( M,\text{ }u^{\theta }\right) $ is the set of strategy
profiles that survive one round of deletion of strictly dominated
strategies, and $UD^{\infty }\left( M,\text{ }u^{\theta }\right) $ is the
set of strategy profiles that survive iterative deletion of strictly
dominated strategies.

For any $\theta \in \Theta ^{\ast }$ and any $\Omega \subseteq \cup _{%
\widetilde{\theta }\in \Theta ^{\ast }}\Omega ^{\left[ \widetilde{\theta },%
\text{ }%
\mathbb{R}
\right] }$, define 
\begin{equation*}
\Omega _{\theta }\equiv \Omega \cap \Omega ^{\left[ \theta ,\text{ }%
\mathbb{R}
\right] }\text{.}
\end{equation*}%
A cardinal implementation problem is a tuple $\left[ \Theta \text{, \ }%
\Omega \text{, }f:\Theta \longrightarrow Z\right] $ such that 
\begin{equation*}
\Theta \subseteq \Theta ^{\ast }\text{, and }\Omega \subseteq \cup _{\theta
\in \Theta }\Omega ^{\left[ \theta ,\text{ }%
\mathbb{R}
\right] }\text{, and }\Omega _{\theta }\neq \varnothing \text{, }\forall
\theta \in \Theta \text{,}
\end{equation*}%
where $\Theta $ and $\Omega $\ are the sets of \emph{legitimate} ordinal and
cardinal states in the implementation problem, respectively, and $f$ is the
targeted social choice function (SCF).

\begin{define}
\label{def:implementation:UD:cardinal}In a cardinal implementation problem $%
\left[ \Theta \text{, \ }\Omega \text{, }f:\Theta \longrightarrow Z\right] $%
, $f$ is $UD$-implemented by a mechanism $M=\langle S,\ g\rangle $ if%
\begin{equation*}
g\left[ UD\left( M,\text{ }u^{\theta }\right) \right] =\left\{ f\left(
\theta \right) \right\} \text{, }\forall \theta \in \Theta \text{, }\forall
u^{\theta }\in \Omega _{\theta }\text{.}
\end{equation*}%
Furthermore, $f$ is $UD^{\infty }$-implemented by a mechanism $M=\langle S,\
g\rangle $ if%
\begin{equation*}
g\left[ UD^{\infty }\left( M,\text{ }u^{\theta }\right) \right] =\left\{
f\left( \theta \right) \right\} \text{, }\forall \theta \in \Theta \text{, }%
\forall u^{\theta }\in \Omega _{\theta }\text{.}
\end{equation*}
\end{define}

In particular, a cardinal implementation problem $\left[ \Theta \text{, \ }%
\Omega \text{, }f:\Theta \longrightarrow Z\right] $ is ordinal if and only
if $\Omega =\cup _{\theta \in \Theta }\Omega ^{\left[ \theta ,\text{ }%
\mathbb{R}
\right] }$. That is, an ordinal implementation problem considers all
possible cardinal representations of the ordinal states, while a cardinal
implementation problem may not. In order to make our results as strong as
possible, we provide impossibility results in cardinal implementation
problems, and provide possibility results in ordinal implementation problems.

\section{Review of \textbf{B{\"{o}}rgers (1995)}}

\label{sec:Borgers}

In a cardinal implementation problem $\left[ \Theta \text{, \ }\Omega \text{%
, }f:\Theta \longrightarrow Z\right] $, agent $i\in \mathcal{I}$ is a
dictator if and only if%
\begin{equation*}
f\left( \theta \right) \succ _{i}^{\theta }z\text{, }\forall z\in Z\diagdown
\left\{ f\left( \theta \right) \right\} \text{, }\forall \theta \in \Theta 
\text{,}
\end{equation*}%
and $f$ is dictatorial if a dictator exists. Furthermore, a mechanism $%
M=\langle S,\ g:S\longrightarrow \triangle \left( Z\right) \rangle $ is
deterministic if and only if $g\left( S\right) \subseteq Z$.

Consider two preference domains: the unanimity domain (denoted by $\Theta ^{%
\text{una}}$), and the strict-preference domain (denoted by $\Theta ^{\text{%
strict}}$):%
\begin{eqnarray*}
\Theta ^{\text{una}} &\equiv &\left\{ \theta \equiv \left( \succeq
_{i}^{\theta }\right) _{i\in \mathcal{I}}\in \Theta ^{\ast }:z\succeq
_{i}^{\theta }z^{\prime }\Longleftrightarrow z\succeq _{j}^{\theta
}z^{\prime }\text{, }\forall \left( z,z^{\prime },i,j\right) \in Z\times
Z\times \mathcal{I}\times \mathcal{I}\right\} \text{,} \\
\Theta ^{\text{strict}} &\equiv &\left\{ \theta \equiv \left( \succeq
_{i}^{\theta }\right) _{i\in \mathcal{I}}\in \Theta ^{\ast }:\left[ z\succeq
_{i}^{\theta }z^{\prime }\text{ and }z^{\prime }\succeq _{i}^{\theta }z%
\right] \Longrightarrow z=z^{\prime }\text{, }\forall \left( z,z^{\prime
},i\right) \in Z\times Z\times \mathcal{I}\right\} \text{.}
\end{eqnarray*}%
$\Theta ^{\text{una}}$ consists of all ordinal states in which all agents
rank outcomes identically, and $\Theta ^{\text{strict}}$ consists of all
ordinal states without non-trivial indifference. With mild domain
restrictions, \cite{tb95} proves the following impossibility result.

\begin{theo}[\protect\cite{tb95}]
\label{thm:borgers:deterministic}In any cardinal implementation problem $%
\left[ \Theta \text{, \ }\Omega \text{, }f:\Theta \longrightarrow Z\right] $
such that $f$ is surjective and $\Theta ^{\text{una}}\cap \Theta ^{\text{%
strict}}\subseteq \Theta \subseteq \Theta ^{\text{strict}}$, the following
three statements are equivalent:

(i) $f$ can be $UD$-implemented by a deterministic mechanism;

(ii) $f$ can be $UD^{\infty }$-implemented by a deterministic mechanism;

(iii) $f$ is dictatorial.
\end{theo}

Theorem \ref{thm:borgers:deterministic} establishes that only dictatorial
SCFs can be $UD$-implemented or $UD^{\infty }$-implemented by a
deterministic mechanism. As \cite{tb95} observes, the argument exploits a
discrete feature of deterministic mechanisms. With stochastic mechanisms,
this discreteness disappears. Thus, \cite{tb95} raises an open question:
does Theorem \ref{thm:borgers:deterministic} still hold when
\textquotedblleft deterministic mechanism\textquotedblright\ is replaced by
\textquotedblleft stochastic mechanism\textquotedblright ? We aim to answer
this question in this paper.

\section{$UD$-implementation}

\label{sec:impossibility:UD}

We first focus on $UD$-implementation, and show \cite{tb95}'s impossibility
result extends to stochastic mechanisms.

\begin{theo}
\label{thm:UD:finite}In any cardinal implementation problem $\left[ \Theta 
\text{, \ }\Omega \text{, }f:\Theta \longrightarrow Z\right] $ such that $f$
is surjective and $\Theta ^{\text{una}}\cap \Theta ^{\text{strict}}\subseteq
\Theta \subseteq \Theta ^{\text{strict}}$, $f$ can be $UD$-implemented by a
stochastic mechanism if and only if $f$ is dictatorial.
\end{theo}

For any $i\in \mathcal{I}$, define the $i$-dictatorial mechanism as: let $i$
name her top outcome, and we implement it. Clearly, if $f$ is dictatorial
with $i$ being a dictator, the $i$-dictatorial mechanism $UD$-implements $f$%
, which proves the "if" part of Theorem \ref{thm:UD:finite}.

In order to prove the "only if" part of Theorem \ref{thm:UD:finite}, we
develop a few tools. For any mechanism $M=\left\langle S,\text{ }%
g\right\rangle $ and any $i\in \mathcal{I}$, define%
\begin{equation}
S_{i}^{z}\equiv \left\{ s_{i}\in S_{i}:\exists s_{-i}\in S_{-i}\text{, }%
g\left( s_{i},s_{-i}\right) =z\right\} \text{, }\forall \left( i,z\right)
\in \mathcal{I}\times Z\text{.}  \label{kkt1}
\end{equation}

\begin{lemma}
\label{lem:Z:2:surjective}In a cardinal implementation problem $\left[
\Theta \text{, \ }\Omega \text{, }f:\Theta \longrightarrow Z\right] $ such
that $f$ is surjective and $\Theta ^{\text{una}}\cap \Theta ^{\text{strict}%
}\subseteq \Theta \subseteq \Theta ^{\text{strict}}$, if $f$ is $UD$%
-implemented by a stochastic mechanism $M=\left\langle S,\text{ }%
g\right\rangle $, we have%
\begin{equation*}
g\left[ \times _{i\in \mathcal{I}}S_{i}^{z}\right] =\left\{ z\right\} \text{%
, }\forall z\in Z\text{.}
\end{equation*}
\end{lemma}

\begin{lemma}
\label{lem:Z:2:UD:una}In a cardinal implementation problem $\left[ \Theta 
\text{, \ }\Omega \text{, }f:\Theta \longrightarrow Z\right] $ such that $f$
is surjective and $\Theta ^{\text{una}}\cap \Theta ^{\text{strict}}\subseteq
\Theta \subseteq \Theta ^{\text{strict}}$, if $f$ is $UD$-implemented by a
stochastic mechanism $M=\left\langle S,\text{ }g\right\rangle $, we have%
\begin{equation*}
\left( 
\begin{array}{c}
z\neq z^{\prime }\text{ and} \\ 
S_{i}^{z}\subseteq S_{i}^{z^{\prime }}%
\end{array}%
\right) \Longrightarrow S_{i}^{z}=S_{i}^{z^{\prime }}=S_{i}\text{, }\forall
\left( i,z,z^{\prime }\right) \in \mathcal{I}\times Z\times Z\text{.}
\end{equation*}
\end{lemma}

The proofs of Lemmas \ref{lem:Z:2:surjective} and \ref{lem:Z:2:UD:una} are
relegated to Sections \ref{sec:lem:Z:2:surjective} and \ref%
{sec:lem:Z:2:UD:una}, respectively.

\noindent \textbf{Proof of the "only if" part of Theorem \ref{thm:UD:finite}:%
} Suppose $f$ is $UD$-implemented by a stochastic mechanism $M=\left\langle
S,\text{ }g\right\rangle $, and we prove by contradiction that $f$ is
dictatorial. Suppose not. Thus, for every $i\in \mathcal{I}$, there exists $%
\left( \theta ,z\right) \in \Theta \times Z$ such that $z$ is agent $i$'s
top outcome at $\theta $, but $z\neq f\left( \theta \right) $. Thus, for
each $s_{i}\in S_{i}^{z}$, there exists $s_{-i}\in S_{-i}$ such that $%
g\left( s_{i},s_{-i}\right) =z$, which achieves $i$'s maximal utility at $%
\theta $. As a result,%
\begin{equation}
S_{i}^{z}\subseteq UD_{i}\left( M,\text{ }u^{\theta }\right) \text{, }%
\forall u^{\theta }\in \Omega _{\theta }\text{.}  \label{tkk1}
\end{equation}%
Since $f$ is $UD$-implemented by $M=\left\langle S,\text{ }g\right\rangle $,
we have $g\left[ UD\left( M,\text{ }u^{\theta }\right) \right] =\left\{
f\left( \theta \right) \right\} $. By the definition of $S_{i}^{f\left(
\theta \right) }$ (i.e., (\ref{kkt1})), we have 
\begin{equation}
UD_{i}\left( M,\text{ }u^{\theta }\right) \subseteq S_{i}^{f\left( \theta
\right) }\text{, }\forall u^{\theta }\in \Omega _{\theta }\text{.}
\label{tkk2}
\end{equation}%
(\ref{tkk1}) and (\ref{tkk2}) imply $S_{i}^{z}\subseteq S_{i}^{f\left(
\theta \right) }$. Thus, Lemma \ref{lem:Z:2:UD:una} implies%
\begin{equation*}
S_{i}^{f\left( \theta \right) }=S_{i}\text{, }\forall i\in \mathcal{I}\text{.%
}
\end{equation*}%
This further implies%
\begin{equation*}
S_{i}^{\widetilde{z}}\subseteq S_{i}=S_{i}^{f\left( \theta \right) }\text{, }%
\forall \left( i,\widetilde{z}\right) \in \mathcal{I}\times Z\text{,}
\end{equation*}%
which, together with Lemma \ref{lem:Z:2:UD:una}, implies%
\begin{equation}
S_{i}^{\widetilde{z}}=S_{i}\text{, }\forall \left( i,\widetilde{z}\right)
\in \mathcal{I}\times Z\text{.}  \label{tkk3}
\end{equation}%
Pick any $z^{\prime }\neq z^{\prime \prime }$. (\ref{tkk3})\ and Lemma \ref%
{lem:Z:2:surjective} imply%
\begin{equation*}
\left\{ z^{\prime }\right\} =g\left[ \times _{i\in \mathcal{I}%
}S_{i}^{z^{\prime }}\right] =g\left[ S\right] =g\left[ \times _{i\in 
\mathcal{I}}S_{i}^{z^{\prime \prime }}\right] =\left\{ z^{\prime \prime
}\right\} \text{,}
\end{equation*}%
contradicting $z^{\prime }\neq z^{\prime \prime }$.$\blacksquare $

The argument in the proof of Theorem \ref{thm:UD:finite} is very general,
and can be extended beyond expected utility. The same proof still holds, as
long as two assumptions are maintained: (1) agents' preferences on $\Delta
(Z)$ are transitive, and (2) an agent's best and worst outcomes in $Z$
remain her best and worst outcomes in $\Delta (Z)$, respectively.

\subsection{Proof of Lemma \protect\ref{lem:Z:2:surjective}}

\label{sec:lem:Z:2:surjective}

Suppose $f$ is $UD$-implemented by $M=\left\langle S,\text{ }g\right\rangle $%
. Consider any $z\in Z$. Since $f$ is surjective, there exists $\theta \in
\Theta $ such that $f\left( \theta \right) =z$, and%
\begin{equation*}
g\left[ UD\left( M,\text{ }u^{\theta }\right) \right] =\left\{ z\right\} 
\text{, }\forall u^{\theta }\in \Omega _{\theta }\text{.}
\end{equation*}%
Therefore, $\times _{i\in \mathcal{I}}S_{i}^{z}\neq \varnothing $.

Pick any $\theta ^{\prime }\in \Theta ^{\text{una}}\cap \Theta ^{\text{strict%
}}$ such that $z$ is the top outcome of all agents. For each agent $i\in 
\mathcal{I}$ and each $s_{i}\in S_{i}^{z}$, there exists $s_{-i}\in S_{-i}$
such that $g\left( s_{i},s_{-i}\right) =z$, which achieves $i$'s maximal
utility at $\theta ^{\prime }$. As a result,%
\begin{equation*}
S_{i}^{z}\subseteq UD_{i}\left( M,\text{ }u^{\theta ^{\prime }}\right) \text{%
, }\forall i\in \mathcal{I}\text{, }\forall u^{\theta ^{\prime }}\in \Omega
_{\theta ^{\prime }}\text{.}
\end{equation*}%
Since $f$ is $UD$-implemented by $M=\left\langle S,\text{ }g\right\rangle $,%
\begin{equation}
g\left[ \times _{i\in \mathcal{I}}S_{i}^{z}\right] \subseteq g\left[
UD\left( M,\text{ }u^{\theta ^{\prime }}\right) \right] =\left\{ f\left(
\theta ^{\prime }\right) \right\} \text{, }\forall u^{\theta ^{\prime }}\in
\Omega _{\theta ^{\prime }}\text{.}  \label{kff1}
\end{equation}%
Since there exists $s\in \times _{i\in \mathcal{I}}S_{i}^{z}$ such that $%
g\left( s\right) =z$, we have%
\begin{equation}
z\in g\left[ \times _{i\in \mathcal{I}}S_{i}^{z}\right] \text{.}
\label{kff1b}
\end{equation}%
(\ref{kff1}) and (\ref{kff1b}) imply $z=f\left( \theta ^{\prime }\right) $,
and hence, $g\left[ \times _{i\in \mathcal{I}}S_{i}^{z}\right] =\left\{
z\right\} $.$\blacksquare $

\subsection{Proof of Lemma \protect\ref{lem:Z:2:UD:una}}

\label{sec:lem:Z:2:UD:una}

Suppose $f$ is $UD$-implemented by $M=\left\langle S,\text{ }g\right\rangle $%
. Fix any $\left( i,z,z^{\prime }\right) \in \mathcal{I}\times Z\times Z$
such that $z\neq z^{\prime }$. Suppose 
\begin{equation}
S_{i}^{z}\subseteq S_{i}^{z^{\prime }}\text{,}  \label{tkk5}
\end{equation}%
and we aim to prove $S_{i}^{z}=S_{i}^{z^{\prime }}=S_{i}$. Pick any $%
\widehat{\theta }\in \Theta ^{\text{una}}\cap \Theta ^{\text{strict}}$ at
which $z$ and $z^{\prime }$ are the best and the worst outcome of all
agents, respectively. Pick any $u^{\widehat{\theta }}\in \Omega _{\widehat{%
\theta }}$.

For each $s_{j}\in S_{j}^{z}$, there exists $s_{-j}\in S_{-j}$ such that $%
g\left( s_{j},s_{-j}\right) =z$, which achieves $j$'s maximal utility at $u^{%
\widehat{\theta }}$. As a result,%
\begin{equation}
\times _{j\in \mathcal{I}}S_{j}^{z}\subseteq UD\left( M,\text{ }u^{\widehat{%
\theta }}\right) \text{.}  \label{tkk4c}
\end{equation}%
Since $f$ is $UD$-implemented by $M=\left\langle S,\text{ }g\right\rangle $, 
\begin{equation}
g\left[ \times _{j\in \mathcal{I}}S_{j}^{z}\right] \subseteq g\left[
UD\left( M,\text{ }u^{\widehat{\theta }}\right) \right] =\left\{ f\left( 
\widehat{\theta }\right) \right\} \text{.}  \label{tkk4}
\end{equation}%
Since there exists $s\in \times _{j\in \mathcal{I}}S_{j}^{z}$ such that $%
g\left( s\right) =z$, we have%
\begin{equation}
z\in g\left[ \times _{j\in \mathcal{I}}S_{j}^{z}\right] \text{.}
\label{tkk4k}
\end{equation}
(\ref{tkk4}) and (\ref{tkk4k}) imply $z=f\left( \widehat{\theta }\right) $,
and hence, $g\left[ UD\left( M,\text{ }u^{\widehat{\theta }}\right) \right]
=\left\{ z\right\} $, which, together with the definition of $S_{j}^{z}$
(i.e., (\ref{kkt1})), implies%
\begin{equation}
UD\left( M,\text{ }u^{\widehat{\theta }}\right) \subseteq \times _{j\in 
\mathcal{I}}S_{j}^{z}\text{.}  \label{tkk4cc}
\end{equation}
(\ref{tkk4c}) and (\ref{tkk4cc}) imply 
\begin{equation}
UD\left( M,\text{ }u^{\widehat{\theta }}\right) =\times _{j\in \mathcal{I}%
}S_{j}^{z}\text{.}  \label{tkk4b}
\end{equation}

Pick any $s_{-i}^{\ast }\in \times _{j\in \mathcal{I}\diagdown \left\{
i\right\} }S_{j}^{z^{\prime }}$. Since $M=\left\langle S,\text{ }%
g\right\rangle $ is a finite-action mechanism, agent $i$ has a best reply to 
$s_{-i}^{\ast }$ at $u^{\widehat{\theta }}$, and we denote this best reply
by $\widehat{s}_{i}$, i.e.,%
\begin{equation}
U_{i}^{\widehat{\theta }}\left[ g\left( \widehat{s}_{i},s_{-i}^{\ast
}\right) \right] \geq U_{i}^{\widehat{\theta }}\left[ g\left( s_{i}^{\prime
},s_{-i}^{\ast }\right) \right] \text{, }\forall s_{i}^{\prime }\in S_{i}%
\text{.}  \label{raa4}
\end{equation}%
As a result, $\widehat{s}_{i}$ is not strictly dominated at $u^{\widehat{%
\theta }}$, i.e., $\widehat{s}_{i}\in UD_{i}\left( M,\text{ }u^{\widehat{%
\theta }}\right) $. By (\ref{tkk4b}), we have $\widehat{s}_{i}\in S_{i}^{z}$%
, which, together with (\ref{tkk5}), implies%
\begin{equation*}
\widehat{s}_{i}\in S_{i}^{z}\subseteq S_{i}^{z^{\prime }}\text{.}
\end{equation*}%
Recall $s_{-i}^{\ast }\in \times _{j\in \mathcal{I}\diagdown \left\{
i\right\} }S_{j}^{z^{\prime }}$. Lemma \ref{lem:Z:2:surjective} implies $%
g\left( \widehat{s}_{i},s_{-i}^{\ast }\right) =z^{\prime }$. Recall $%
z^{\prime }$ is the worst outcome of all agents at $u^{\widehat{\theta }}$.
We thus have 
\begin{equation}
U_{i}^{\widehat{\theta }}\left[ g\left( s_{i}^{\prime \prime },s_{-i}^{\ast
}\right) \right] \geq U_{i}^{\widehat{\theta }}\left[ z^{\prime }\right]
=U_{i}^{\widehat{\theta }}\left[ g\left( \widehat{s}_{i},s_{-i}^{\ast
}\right) \right] \text{, }\forall s_{i}^{\prime \prime }\in S_{i}\text{.}
\label{raa4b}
\end{equation}%
(\ref{raa4}) and (\ref{raa4b}) imply every $s_{i}^{\prime \prime }\in S_{i}$
is a best reply to $s_{-i}^{\ast }$ at $u^{\widehat{\theta }}$, and hence
not strictly dominated at $u^{\widehat{\theta }}$, i.e., $S_{i}\subseteq
UD_{i}\left( M,\text{ }u^{\widehat{\theta }}\right) $. By (\ref{tkk4b}), we
have $S_{i}\subseteq S_{i}^{z}$, which, together with (\ref{tkk5}), implies%
\begin{equation*}
S_{i}\subseteq S_{i}^{z}\subseteq S_{i}^{z^{\prime }}\text{.}
\end{equation*}%
Therefore, $S_{i}^{z}=S_{i}^{z^{\prime }}=S_{i}$.$\blacksquare $

\section{$UD^{\infty }$-implementation}

\label{sec:impossibility:UD-infinite}

We study $UD^{\infty }$-implementation in this section. When there are two
outcomes only, an impossibility result is provided in Section \ref%
{sec:impossibility:2:UD-infinite}. When there are three or more outcomes,
possibility results are provided in Section \ref{sec:possibility:finite}.

\subsection{Two-outcome environments}

\label{sec:impossibility:2:UD-infinite}

Given $\left\vert Z\right\vert =2$, \cite{tb95}'s impossibility result on $%
UD^{\infty }$-implementation extends to stochastic mechanisms.

\begin{theo}
\label{thm:UD2:2-outcomes}Suppose $\left\vert Z\right\vert =2$. In any
cardinal implementation problem $\left[ \Theta \text{, \ }\Omega \text{, }%
f:\Theta \longrightarrow Z\right] $ such that $f$ is surjective and $\Theta
^{\text{una}}\cap \Theta ^{\text{strict}}\subseteq \Theta \subseteq \Theta ^{%
\text{strict}}$, $f$ can be $UD^{\infty }$-implemented by a stochastic
mechanism if and only if $f$ is dictatorial.
\end{theo}

Clearly, if $f$ is dictatorial with $i$ being a dictator, the $i$%
-dictatorial mechanism $UD^{\infty }$-implements $f$, which proves the "if"
part of Theorem \ref{thm:UD2:2-outcomes}. We need the following two lemmas
to prove the "only if" part.

\begin{lemma}
\label{lem:Z:2:UD-infinite}In a cardinal implementation problem $\left[
\Theta \text{, \ }\Omega \text{, }f:\Theta \longrightarrow Z\right] $ such
that $f$ is surjective and $\Theta ^{\text{una}}\cap \Theta ^{\text{strict}%
}\subseteq \Theta \subseteq \Theta ^{\text{strict}}$, if $f$ is $UD^{\infty
} $-implemented by a stochastic mechanism $M=\left\langle S,\text{ }%
g\right\rangle $, we have%
\begin{equation*}
g\left[ \times _{i\in \mathcal{I}}S_{i}^{z}\right] =\left\{ z\right\} \text{%
, }\forall z\in Z\text{.}
\end{equation*}
\end{lemma}

For $UD^{\infty }$-implementation, Lemma \ref{lem:Z:2:UD-infinite} is the
counterpart of Lemma \ref{lem:Z:2:surjective} for $UD$-implementation. The
proof of Lemma \ref{lem:Z:2:UD-infinite} is relegated to Section \ref%
{sec:lem:Z:2:UD-infinite}.

\begin{lemma}
\label{lem:Z:2:UD-infinite:2}Suppose $\left\vert Z\right\vert =2$. In any
cardinal implementation problem $\left[ \Theta \text{, \ }\Omega \text{, }%
f:\Theta \longrightarrow Z\right] $ such that $f$ is surjective, if $f$ is $%
UD^{\infty }$-implemented by a stochastic mechanism $M=\left\langle S,\text{ 
}g\right\rangle $ and agent $i\in \mathcal{I}$ is not a dictator, we have%
\begin{equation*}
\cap _{z\in Z}S_{i}^{z}\neq \varnothing \text{.}
\end{equation*}
\end{lemma}

Lemma \ref{lem:Z:2:UD-infinite:2} says that if agent $i$ is not a dictator,
there exists a strategy $s_{i}\in \cap _{z\in Z}S_{i}^{z}$ that does not pin
down the outcome chosen by $M=\left\langle S,\text{ }g\right\rangle $, i.e.,
depending on agent $-i$'s strategies, the mechanism may choose any outcome
in $Z$. The proof of Lemma \ref{lem:Z:2:UD-infinite:2} is relegated to
Section \ref{sec:lem:Z:2:UD-infinite:2}.

\noindent \textbf{Proof of the "only if" part of Theorem \ref%
{thm:UD2:2-outcomes}:} Suppose $\left\vert Z\right\vert =2$, and $f$ is $%
UD^{\infty }$-implemented by $M=\left\langle S,\text{ }g\right\rangle $. We
prove by contradiction that $f$ is dictatorial. Suppose not.

Fix any $z^{\prime },z^{\prime \prime }\in Z$ such that $z^{\prime }\neq
z^{\prime \prime }$. Since $f$ is $UD^{\infty }$-implemented by $%
M=\left\langle S,\text{ }g\right\rangle $, Lemma \ref{lem:Z:2:UD-infinite}\
implies%
\begin{equation}
g\left[ \times _{i\in \mathcal{I}}S_{i}^{z^{\prime }}\right] =\left\{
z^{\prime }\right\} \text{ and }g\left[ \times _{i\in \mathcal{I}%
}S_{i}^{z^{\prime \prime }}\right] =\left\{ z^{\prime \prime }\right\} \text{%
.}  \label{kkt9}
\end{equation}%
Since $f$ is not dictatorial, Lemma \ref{lem:Z:2:UD-infinite:2} implies $%
\times _{i\in \mathcal{I}}\left[ \cap _{z\in Z}S_{i}^{z}\right] \neq
\varnothing $. As a result, (\ref{kkt9})\ implies%
\begin{equation*}
\left\{ z^{\prime }\right\} =g\left[ \times _{i\in \mathcal{I}%
}S_{i}^{z^{\prime }}\right] \supseteq g\left( \times _{i\in \mathcal{I}}%
\left[ \cap _{z\in Z}S_{i}^{z}\right] \right) \subseteq g\left[ \times
_{i\in \mathcal{I}}S_{i}^{z^{\prime \prime }}\right] =\left\{ z^{\prime
\prime }\right\} \text{,}
\end{equation*}%
contradicting $z^{\prime }\neq z^{\prime \prime }$.$\blacksquare $

\subsubsection{Proof of Lemma \protect\ref{lem:Z:2:UD-infinite}}

\label{sec:lem:Z:2:UD-infinite}

Suppose $f$ is $UD^{\infty }$-implemented by $M=\left\langle S,\text{ }%
g\right\rangle $. Consider any $z\in Z$. Since $f$ is surjective, there
exists $\theta \in \Theta $ such that $f\left( \theta \right) =z$, and%
\begin{equation*}
g\left[ UD\left( M,\text{ }u^{\theta }\right) \right] =\left\{ z\right\} 
\text{, }\forall u^{\theta }\in \Omega _{\theta }\text{.}
\end{equation*}%
Therefore, $\times _{i\in \mathcal{I}}S_{i}^{z}\neq \varnothing $.

Pick any $\theta ^{\prime }\in \Theta ^{\text{una}}\cap \Theta ^{\text{strict%
}}$ such that $z$ is the top outcome of all agents. For each agent $i\in 
\mathcal{I}$ and each $s_{i}\in S_{i}^{z}$, there exists $s_{-i}\in
S_{-i}^{z}$ such that $g\left( s_{i},s_{-i}\right) =z$, which achieves $i$'s
maximal utility at $\theta ^{\prime }$. As a result,%
\begin{equation*}
S_{i}^{z}\subseteq UD_{i}^{1}\left( M,\text{ }u^{\theta ^{\prime }}\right) 
\text{, }\forall i\in \mathcal{I}\text{, }\forall u^{\theta ^{\prime }}\in
\Omega _{\theta ^{\prime }}\text{.}
\end{equation*}%
Inductively, the same argument shows%
\begin{equation*}
S_{i}^{z}\subseteq UD_{i}^{k}\left( M,\text{ }u^{\theta ^{\prime }}\right) 
\text{, }\forall i\in \mathcal{I}\text{, }\forall u^{\theta ^{\prime }}\in
\Omega _{\theta ^{\prime }}\text{, }\forall k\in \mathbb{N}\text{,}
\end{equation*}%
and as a result, $\times _{i\in \mathcal{I}}S_{i}^{z}\subseteq UD^{\infty
}\left( M,\text{ }u^{\theta ^{\prime }}\right) $. Since $f$ is $UD^{\infty }$%
-implemented by $M=\left\langle S,\text{ }g\right\rangle $,%
\begin{equation}
g\left[ \times _{i\in \mathcal{I}}S_{i}^{z}\right] \subseteq g\left[
UD^{\infty }\left( M,\text{ }u^{\theta ^{\prime }}\right) \right] =\left\{
f\left( \theta ^{\prime }\right) \right\} \text{, }\forall u^{\theta
^{\prime }}\in \Omega _{\theta ^{\prime }}\text{.}  \label{bnn1}
\end{equation}%
Since there exists $s\in \times _{i\in \mathcal{I}}S_{i}^{z}$ such that $%
g\left( s\right) =z$, we have%
\begin{equation}
z\in g\left[ \times _{i\in \mathcal{I}}S_{i}^{z}\right] \text{.}
\label{bnn2}
\end{equation}%
(\ref{bnn1}) and (\ref{bnn2}) imply $z=f\left( \theta ^{\prime }\right) $,
and hence, $g\left[ \times _{i\in \mathcal{I}}S_{i}^{z}\right] =\left\{
z\right\} $.$\blacksquare $

\subsubsection{Proof of Lemma \protect\ref{lem:Z:2:UD-infinite:2}}

\label{sec:lem:Z:2:UD-infinite:2}

Given a mechanism $M=\left\langle S,\text{ }g\right\rangle $, we say $%
\widetilde{S}\equiv \times _{j\in I}\widetilde{S}_{j}\subseteq S$ satisfies
the non-domination property at a cardinal state $u^{\theta }\equiv \left(
u_{j}^{\theta }\right) _{j\in \mathcal{I}}$ if and only if for every $i\in 
\mathcal{I}$ and every $s_{i}\in \widetilde{S}_{i}$, $s_{i}$ is not strictly
dominated on $\widetilde{S}$, or equivalently, 
\begin{equation*}
\forall i\in \mathcal{I}\text{, }\forall \left( s_{i},s_{i}^{\prime }\right)
\in \widetilde{S}_{i}\times S_{i}\text{, }\exists s_{-i}\in \widetilde{S}%
_{-i}\text{, }U_{i}^{\theta }\left[ g\left( s_{i},s_{-i}\right) \right] \geq
U_{i}^{\theta }\left[ g\left( s_{i}^{\prime },s_{-i}\right) \right] \text{.}
\end{equation*}%
It is straightforward to show that $UD^{\infty }\left( M,\text{ }u^{\theta
}\right) $ is the largest set that satisfies the non-domination property.%
\footnote{%
The non-domination property for iterative deletion of strictly dominated
strategies is analogous to the best-reply property for iterative deletion of
never-best strategies. Precisely, iterative deletion of never-best
strategies is the largest set that satisfies the best-reply property, while
iterative deletion of strictly dominated strategies is the largest set that
satisfies the non-domination property.} It will be clear that the
non-domination property is the driving force for Lemma \ref%
{lem:Z:2:UD-infinite:2}.

\noindent \textbf{Proof of Lemma \ref{lem:Z:2:UD-infinite:2}:} Since $f$ is $%
UD^{\infty }$-implemented by $M=\left\langle S,\text{ }g\right\rangle $ and $%
f$ is surjective, we have 
\begin{equation}
S_{i}^{z}\neq \varnothing \text{, }\forall z\in Z\text{.}  \label{kkt11}
\end{equation}%
Since agent $i$ is not a dictator, there exists $\left( \theta ,z^{\prime
},z^{\prime \prime }\right) $ such that $z^{\prime \prime }\succ
_{i}^{\theta }z^{\prime }$ and $f\left( \theta \right) =z^{\prime }$. Given $%
\left\vert Z\right\vert =2$, $z^{\prime }$ is the worst outcome for $i$ at $%
\theta $.

Pick any $u^{\theta }\in \Omega _{\theta }$. Fix any $\widehat{s}_{i}\in
UD_{i}^{\infty }\left( M,\text{ }u^{\theta }\right) $. Since $UD^{\infty
}\left( M,\text{ }u^{\theta }\right) $ satisfies the non-domination
property, we have%
\begin{equation}
\forall s_{i}\in S_{i}\text{, }\exists \widehat{s}_{-i}\in UD_{-i}^{\infty
}\left( M,\text{ }u^{\theta }\right) \text{, \ }u_{i}^{\theta }\left(
z^{\prime }\right) =U_{i}^{\theta }\left( g\left[ \widehat{s}_{i},\widehat{s}%
_{-i}\right] \right) \geq U_{i}^{\theta }\left( g\left[ s_{i},\widehat{s}%
_{-i}\right] \right) \text{.}  \label{kkt10}
\end{equation}%
Since $z^{\prime }$ is the unique worst outcome of agent $i$ in $\Delta (Z)$
at $u^{\theta }$, (\ref{kkt10}) implies%
\begin{equation*}
\forall s_{i}\in S_{i}\text{, }\exists \widehat{s}_{-i}\in UD_{-i}^{\infty
}\left( M,\text{ }u^{\theta }\right) \text{, \ }g\left[ s_{i},\widehat{s}%
_{-i}\right] =z^{\prime }\text{.}
\end{equation*}%
By the definition of $S_{i}^{z^{\prime }}$ (i.e., (\ref{kkt1})), we have $%
S_{i}^{z^{\prime }}=S_{i}$. Given $\left\vert Z\right\vert =2$, we thus have%
\begin{equation*}
\cap _{z\in Z}S_{i}^{z}=S_{i}^{z^{\prime }}\cap S_{i}^{z^{\prime \prime
}}=S_{i}\cap S_{i}^{z^{\prime \prime }}=S_{i}^{z^{\prime \prime }}\neq
\varnothing \text{,}
\end{equation*}%
where the inequality follows from (\ref{kkt11}).$\blacksquare $

\subsection{Three or more outcomes}

\label{sec:possibility:finite}

We provide possibility results for three or more outcomes. To make our
construction as simple as possible, we first consider the case of two agents
and three outcomes. The idea will be extended to general environments in
Section \ref{sec:UD2:3-outcomes}. Consider:%
\begin{gather*}
\mathcal{I}=\left\{ i_{1},i_{2}\right\} \text{, }Z=\left\{ a,b,c\right\} 
\text{, and state }\widehat{\theta }:\left( 
\begin{array}{c}
\text{agent }i_{1}:\text{ }b\succ _{i_{1}}^{\widehat{\theta }}a\succ
_{i_{1}}^{\widehat{\theta }}c\text{,} \\ 
\\ 
\text{agent }i_{2}:\text{ }c\succ _{i_{2}}^{\widehat{\theta }}a\succ
_{i_{2}}^{\widehat{\theta }}b\text{.}%
\end{array}%
\right) \text{;} \\
\\
\widehat{\Theta }\equiv \left[ \Theta ^{\text{una}}\cap \Theta ^{\text{strict%
}}\right] \cup \left\{ \widehat{\theta }\right\} \subseteq \Theta ^{\text{%
strict}}\text{, and }\widehat{\Omega }\equiv \cup _{\theta \in \widehat{%
\Theta }}\Omega ^{\left[ \theta ,\text{ }%
\mathbb{R}
\right] }\text{.}
\end{gather*}%
For each $\theta \in \Theta ^{\text{una}}\cap \Theta ^{\text{strict}}$, let $%
z^{\theta }$ denote the top-ranked outcome of the agents. Define $\widehat{f}%
:\widehat{\Theta }\longrightarrow Z$ as%
\begin{equation*}
\widehat{f}\left( \theta \right) \equiv \left\{ 
\begin{tabular}{ll}
$z^{\theta }\text{,}$ & if $\theta \in \Theta ^{\text{una}}\cap \Theta ^{%
\text{strict}}$, \\ 
&  \\ 
$a\text{,}$ & if $\theta =\widehat{\theta }$.%
\end{tabular}%
\right.
\end{equation*}%
We consider the ordinal implementation problem $\left[ \widehat{\Theta }%
\text{, \ }\widehat{\Omega }\text{, }\widehat{f}\right] $, and $\widehat{f}$
is not dictatorial, because $\widehat{f}\left( \widehat{\theta }\right) =a$
is not the top-ranked outcome for both agents at $\widehat{\theta }$. We now
define a mechanism $\widehat{M}=\left\langle \widehat{S},\text{ }\widehat{g}%
\right\rangle $ that $UD^{\infty }$-implements $\widehat{f}$.%
\begin{equation*}
\widehat{S}_{i_{1}}=\widehat{S}_{i_{2}}=Z=\left\{ a,b,c\right\} \text{ and }%
\widehat{S}\equiv \widehat{S}_{i_{1}}\times \widehat{S}_{i_{2}}\text{,}
\end{equation*}%
and $\widehat{g}:\widehat{S}\longrightarrow \triangle \left( Z\right) $ is
described by the following table.%
\begin{equation*}
\begin{tabular}{|c|c|c|c|}
\hline
$\widehat{g}\left( \widehat{s}_{i_{1}},\widehat{s}_{i_{2}}\right) =$ & $%
\widehat{s}_{i_{2}}=a$ & $\widehat{s}_{i_{2}}=b$ & $\widehat{s}_{i_{2}}=c$
\\ \hline
$\widehat{s}_{i_{1}}=a$ & $a$ & $\frac{1}{4}a+\frac{3}{4}b$ & $\frac{1}{2}a+%
\frac{1}{2}b$ \\ \hline
$\widehat{s}_{i_{1}}=b$ & $\frac{1}{2}a+\frac{1}{2}c$ & $b$ & $\frac{1}{2}b+%
\frac{1}{2}c$ \\ \hline
$\widehat{s}_{i_{1}}=c$ & $\frac{1}{4}a+\frac{3}{4}c$ & $\frac{1}{2}b+\frac{1%
}{2}c$ & $c$ \\ \hline
\end{tabular}%
\end{equation*}

\begin{lemma}
\label{lem:Z:3:UD-finite}In the ordinal implementation problem $\left[ 
\widehat{\Theta }\text{, \ }\widehat{\Omega }\text{, }\widehat{f}\right] $,
we have%
\begin{equation}
UD_{i}^{k}\left( \widehat{M},\text{ }u^{\theta }\right) =\left\{ \widehat{f}%
\left( \theta \right) \right\} \text{, }\forall i\in \mathcal{I}\text{, }%
\forall \theta \in \widehat{\Theta }\text{, }\forall u^{\theta }\in \widehat{%
\Omega }_{\theta }\text{, }\forall k\geq 2\text{, }  \label{daa1}
\end{equation}%
i.e., $\widehat{f}$ is $UD^{\infty }$-implemented by $\widehat{M}$.
\end{lemma}

\noindent \textbf{Proof of Lemma \ref{lem:Z:3:UD-finite}:} There are seven
ordinal states in $\widehat{\Theta }$, which are listed as follows.%
\begin{equation*}
\widehat{\Theta }=\left\{ \theta ^{\left[ a\succ b\succ c\right] },\theta ^{%
\left[ a\succ c\succ b\right] },\theta ^{\left[ b\succ a\succ c\right]
},\theta ^{\left[ b\succ c\succ a\right] },\theta ^{\left[ c\succ a\succ b%
\right] },\theta ^{\left[ c\succ b\succ a\right] }\right\} \cup \left\{ 
\widehat{\theta }\right\} \text{,}
\end{equation*}%
where $\theta ^{\left[ a\succ b\succ c\right] }$ denote the state at which
all agents share the preference $\left[ a\succ b\succ c\right] $, and
similar notation applies the other states in $\Theta ^{\text{una}}\cap
\Theta ^{\text{strict}}$. We show (\ref{daa1}) by conducting iterative
deletion of strictly dominated strategies as follows.%
\begin{equation*}
\begin{tabular}{|c|c|c|c|c|c|c|c|}
\hline
$\theta =$ & $\widehat{\theta }$ & $\theta ^{\left[ a\succ b\succ c\right] }$
& $\theta ^{\left[ a\succ c\succ b\right] }$ & $\theta ^{\left[ b\succ
a\succ c\right] }$ & $\theta ^{\left[ b\succ c\succ a\right] }$ & $\theta ^{%
\left[ c\succ a\succ b\right] }$ & $\theta ^{\left[ c\succ b\succ a\right] }$
\\ \hline
$UD_{i_{1}}^{1}\left( \widehat{M},\text{ }u^{\theta }\right) =$ & $\left\{
a,b\right\} $ & $\left\{ a\right\} $ & $\left\{ a,c\right\} $ & $\left\{
a,b\right\} $ & $\left\{ b,c\right\} $ & $\left\{ c\right\} $ & $\left\{
c\right\} $ \\ \hline
$UD_{i_{2}}^{1}\left( \widehat{M},\text{ }u^{\theta }\right) =$ & $\left\{
a,c\right\} $ & $\left\{ a,b\right\} $ & $\left\{ a\right\} $ & $\left\{
b\right\} $ & $\left\{ b\right\} $ & $\left\{ a,c\right\} $ & $\left\{
b,c\right\} $ \\ \hline
$\forall i\in \mathcal{I}$, $UD_{i}^{2}\left( \widehat{M},\text{ }u^{\theta
}\right) =$ & $\left\{ a\right\} $ & $\left\{ a\right\} $ & $\left\{
a\right\} $ & $\left\{ b\right\} $ & $\left\{ b\right\} $ & $\left\{
c\right\} $ & $\left\{ c\right\} $ \\ \hline
\end{tabular}%
\end{equation*}%
$\blacksquare $

\subsubsection{A general possibility result}

\label{sec:UD2:3-outcomes}

Lemma \ref{lem:Z:3:UD-finite} provides a possibility result for two agents
and three outcomes. We extend this result to environments with more agents
and outcomes.

For any two distinct agents $i_{1},i_{2}\in \mathcal{I}$ and any $z\in Z$,
define%
\begin{equation*}
\Theta ^{\left( i_{1},i_{2}\right) \text{-}z}\equiv \left\{ \theta \in
\Theta ^{\text{strict}}:%
\begin{tabular}{c}
$\left\vert \left\{ \widetilde{z}\in Z:z\succeq _{i_{1}}^{\theta }\widetilde{%
z}\right\} \right\vert =\left\vert \left\{ \widetilde{z}\in Z:z\succeq
_{i_{2}}^{\theta }\widetilde{z}\right\} \right\vert =\left\vert Z\right\vert
-1$, \\ 
and \\ 
$\left\{ \widetilde{z}\in Z:z\succeq _{i_{1}}^{\theta }\widetilde{z}\right\}
\neq \left\{ \widetilde{z}\in Z:z\succeq _{i_{2}}^{\theta }\widetilde{z}%
\right\} $%
\end{tabular}%
\right\} \text{.}
\end{equation*}%
At any $\theta \in \Theta ^{\left( i_{1},i_{2}\right) \text{-}z}$, $z$ is
ranked second-best for both $i_{1}$ and $i_{2}$, and their top-ranked
outcomes are different.

\begin{theo}
\label{thm:UD2:3-outcome}Consider any $\overline{\theta }\in \Theta ^{\text{%
strict}}$. In any ordinal implementation problem 
\begin{equation*}
\left( \overline{\Theta }=\left\{ \overline{\theta }\right\} \cup \left[
\Theta ^{\text{una}}\cap \Theta ^{\text{strict}}\right] \text{, \ }\Omega
=\cup _{\theta \in \overline{\Theta }}\Omega ^{\left[ \theta ,\text{ }%
\mathbb{R}
\right] }\text{, }f:\overline{\Theta }\longrightarrow Z\right)
\end{equation*}%
such that $f\left( \theta \right) $ is agents' top-ranked outcome for any $%
\theta \in \Theta ^{\text{una}}\cap \Theta ^{\text{strict}}$, $f$ can be $%
UD^{\infty }$-implemented by a stochastic mechanism if there exist two
distinct agents $i_{1},i_{2}\in \mathcal{I}$ such that%
\begin{equation*}
\overline{\theta }\in \Theta ^{\left( i_{1},i_{2}\right) \text{-}f\left( 
\overline{\theta }\right) }\text{.}
\end{equation*}
\end{theo}

When $\left\vert \mathcal{I}\right\vert =2$ and $\left\vert Z\right\vert =3$%
, Theorem \ref{thm:UD2:3-outcome} reduces to Lemma \ref{lem:Z:3:UD-finite},
and Theorem \ref{thm:UD2:3-outcome} extends the result to environments with $%
\left\vert \mathcal{I}\right\vert >2$ or $\left\vert Z\right\vert >3$.

\noindent \textbf{Proof of Theorem \ref{thm:UD2:3-outcome}:} If $\left\vert
Z\right\vert =3$, we follow the same proof of Lemma \ref{lem:Z:3:UD-finite}
to show that $\widehat{M}=\left\langle \widehat{S},\text{ }\widehat{g}%
\right\rangle $ defined above $UD^{\infty }$-implements $f$. In particular,
we invite agents $i_{1}$ and $i_{2}$ only to play this game.

Similarly, if $\left\vert Z\right\vert >3$, we define a mechanism $M^{\ast
}=\left\langle S^{\ast },\text{ }g^{\ast }\right\rangle $ that invites $%
i_{1} $ and $i_{2}$ only to participate. Specifically, since $\overline{%
\theta }\in \Theta ^{\left( i_{1},i_{2}\right) \text{-}f\left( \overline{%
\theta }\right) }$, there exist three distinct outcomes $a,b,c\in Z$ such
that%
\begin{equation*}
\left( 
\begin{array}{c}
f\left( \overline{\theta }\right) =a\text{ is the second-best outcome of
agents }i_{1}\text{ and }i_{2}\text{ at }\overline{\theta }\text{,} \\ 
b\text{ is agent }i_{1}\text{'s top-ranked outcome at }\overline{\theta }%
\text{,} \\ 
c\text{ is agent }i_{2}\text{'s top-ranked outcome at }\overline{\theta }%
\text{.}%
\end{array}%
\right)
\end{equation*}%
Define 
\begin{equation*}
S_{i_{1}}^{\ast }=S_{i_{2}}^{\ast }=Z\text{ and }S^{\ast }\equiv
S_{i_{1}}^{\ast }\times S_{i_{2}}^{\ast }\text{,}
\end{equation*}%
and $g^{\ast }:S^{\ast }\longrightarrow \triangle \left( Z\right) $ is
defined as follows.%
\begin{equation*}
g^{\ast }\left( s_{i_{1}}^{\ast }=z,\text{ }s_{i_{2}}^{\ast }=z\right) =z%
\text{, }\forall z\in Z\text{,}
\end{equation*}%
and with $s_{i_{1}}^{\ast }\neq $ $s_{i_{2}}^{\ast }$, $g^{\ast }\left(
s_{i_{1}}^{\ast },\text{ }s_{i_{2}}^{\ast }\right) $ is described by the
following table:%
\begin{equation*}
\begin{tabular}{|c|c|c|c|}
\hline
given $s_{i_{1}}^{\ast }\neq $ $s_{i_{2}}^{\ast }$, $g^{\ast }\left(
s_{i_{1}}^{\ast },\text{ }s_{i_{2}}^{\ast }\right) =$ & $s_{i_{2}}^{\ast }=a$
& $s_{i_{2}}^{\ast }=c$ & $s_{i_{2}}^{\ast }\in Z\diagdown \left\{
a,c\right\} $ \\ \hline
$s_{i_{1}}^{\ast }=a$ &  & $\frac{3}{4}a+\frac{1}{4}b$ & $\frac{1}{2}a+\frac{%
1}{2}s_{i_{2}}^{\ast }$ \\ \hline
$s_{i_{1}}^{\ast }=b$ & $\frac{3}{4}a+\frac{1}{4}c$ & $\frac{1}{2}a+\frac{1}{%
4}b+\frac{1}{4}c$ & $\frac{1}{2}b+\frac{1}{2}s_{i_{2}}^{\ast }$ \\ \hline
$s_{i_{1}}^{\ast }\in Z\diagdown \left\{ a,b\right\} $ & $\frac{1}{2}%
s_{i_{1}}^{\ast }+\frac{1}{2}a$ & $\frac{1}{2}s_{i_{1}}^{\ast }+\frac{1}{2}c$
& $\frac{1}{2}s_{i_{1}}^{\ast }+\frac{1}{2}s_{i_{2}}^{\ast }$ \\ \hline
\end{tabular}%
\end{equation*}%
we will show%
\begin{equation}
UD_{i}^{k}\left( M^{\ast },\text{ }u^{\theta }\right) =\left\{ f\left(
\theta \right) \right\} \text{, }\forall i\in \left\{ i_{1},i_{2}\right\} 
\text{, }\forall \theta \in \overline{\Theta }\text{, }\forall u^{\theta
}\in \Omega _{\theta }\text{, }\forall k\geq 3\text{,}  \label{daa2}
\end{equation}%
and as a result,%
\begin{equation*}
g\left[ UD^{\infty }\left( M^{\ast },\text{ }u^{\theta }\right) \right]
=f\left( \theta \right) \text{, }\forall \theta \in \overline{\Theta }\text{%
, }\forall u^{\theta }\in \Omega _{\theta }\text{,}
\end{equation*}%
i.e., $f$ is $UD^{\infty }$-implemented by $M^{\ast }$.

We partition $\overline{\Theta }$ into seven groups:%
\begin{gather*}
\text{(i)}\text{: state }\overline{\theta }\text{;} \\
\text{(ii)}\text{: }\theta ^{\left[ a\succ b\succ c\right] }\in \Theta ^{%
\text{una}}\cap \Theta ^{\text{strict}}\text{ at which }a\text{ is
top-ranked, and }b\text{ is strictly preferred to }c\text{;} \\
\text{(iii)}\text{: }\theta ^{\left[ a\succ c\succ b\right] }\in \Theta ^{%
\text{una}}\cap \Theta ^{\text{strict}}\text{ at which }a\text{ is
top-ranked, and }c\text{ is strictly preferred to }b\text{;} \\
\text{(iv)}\text{: }\theta ^{\left[ b\right] }\in \Theta ^{\text{una}}\cap
\Theta ^{\text{strict}}\text{ at which }b\text{ is top-ranked;} \\
\text{(v)}\text{: }\theta ^{\left[ c\right] }\in \Theta ^{\text{una}}\cap
\Theta ^{\text{strict}}\text{ at which }c\text{ is top-ranked;} \\
\text{(vi)}\text{: }\theta ^{\left[ z\succ b\succ c\right] }\in \Theta ^{%
\text{una}}\cap \Theta ^{\text{strict}}\text{ at which }z\left( \neq
a,b,c\right) \text{ is top-ranked, and }b\text{ is strictly preferred to }c%
\text{;} \\
\text{(vii)}\text{: }\theta ^{\left[ z\succ c\succ b\right] }\in \Theta ^{%
\text{una}}\cap \Theta ^{\text{strict}}\text{ at which }z\left( \neq
a,b,c\right) \text{ is top-ranked, and }c\text{ is strictly preferred to }b%
\text{.}
\end{gather*}%
We show (\ref{daa2}) by conducting iterative deletion of strictly dominated
strategies as follows.%
\begin{equation*}
\begin{tabular}{|c|c|c|c|c|c|c|c|}
\hline
$\theta =$ & $\overline{\theta }$ & $\theta ^{\left[ a\succ b\succ c\right]
} $ & $\theta ^{\left[ a\succ c\succ b\right] }$ & $\theta ^{\left[ b\right]
}$ & $\theta ^{\left[ c\right] }$ & 
\begin{tabular}{l}
$\theta ^{\left[ z\succ b\succ c\right] }$ with \\ 
$z\notin \left\{ a,b,c\right\} $%
\end{tabular}
& 
\begin{tabular}{l}
$\theta ^{\left[ z\succ c\succ b\right] }$ with \\ 
$z\notin \left\{ a,b,c\right\} $%
\end{tabular}
\\ \hline
$UD_{i_{1}}^{1}\left( M^{\ast },\text{ }u^{\theta }\right) \subseteq $ & $%
\left\{ a,b\right\} $ & $\left\{ a\right\} $ & $Z$ & $Z$ & $\left\{
c\right\} $ & $Z$ & $\left\{ a,z\right\} $ \\ \hline
$UD_{i_{2}}^{1}\left( M^{\ast },\text{ }u^{\theta }\right) \subseteq $ & $%
\left\{ a,c\right\} $ & $Z$ & $\left\{ a\right\} $ & $\left\{ b\right\} $ & $%
Z$ & $\left\{ a,z\right\} $ & $Z$ \\ \hline
\begin{tabular}{l}
$\forall i\in \left\{ i_{1},i_{2}\right\} $, \\ 
$UD_{i}^{2}\left( M^{\ast },\text{ }u^{\theta }\right) \subseteq $%
\end{tabular}
& $\left\{ a\right\} $ & $\left\{ a\right\} $ & $\left\{ a\right\} $ & $%
\left\{ b\right\} $ & $\left\{ c\right\} $ & $\left\{ a,z\right\} $ & $%
\left\{ a,z\right\} $ \\ \hline
\begin{tabular}{l}
$\forall i\in \left\{ i_{1},i_{2}\right\} $, \\ 
$UD_{i}^{3}\left( M^{\ast },\text{ }u^{\theta }\right) \subseteq $%
\end{tabular}
&  &  &  &  &  & $\left\{ z\right\} $ & $\left\{ z\right\} $ \\ \hline
\end{tabular}%
\end{equation*}%
Here, $UD_{i}^{k}\left( M^{\ast },\text{ }u^{\theta }\right) $ may depend on
the cardinal uitlity, and hence, we only identify a superset of $%
UD_{i}^{k}\left( M^{\ast },\text{ }u^{\theta }\right) $, or equivalently, we
may not eliminate all of the dominated strategies at each round.$%
\blacksquare $

\section{Infinite-action mechanisms}

\label{sec:infinite-action}

We consider infinite-action mechanisms in this section. Finite-action
mechanisms are degenerate infinite-action mechanisms, and hence, given $%
\left\vert Z\right\vert \geq 3$, the possibility result for $UD^{\infty }$%
-implementation (Theorem \ref{thm:UD2:3-outcome}) remains true for
infinite-action mechanisms.

As argued by \cite{bl1994}, $UD^{\infty }$ should be defined by transfinite
deletion of strictly dominated strategies. Given such a definition, it is
easy to show that $UD^{\infty }$ satisfies the non-domination property,
which is the driving force for the impossibility result for $UD^{\infty }$%
-implementation given $\left\vert Z\right\vert =2$ (see Section \ref%
{sec:lem:Z:2:UD-infinite:2}). Therefore, Theorem \ref{thm:UD2:2-outcomes}
remains true for infinite-action mechanisms.

We now prove a possibility result for $UD$-implementation by infinite-action
mechanisms, which is in contrast to impossibility of $UD$-implementation by
finite-action mechanisms (Theorem \ref{thm:UD:finite}). Therefore,
infinite-action mechanisms expand the scope of $UD$-implementation.

\begin{theo}
\label{thm:UD:infinite}Consider any $\theta \in \Theta ^{\text{strict}}$. In
any ordinal implementation problem 
\begin{equation*}
\left[ \Theta =\left( \Theta ^{\text{una}}\cap \Theta ^{\text{strict}%
}\right) \cup \left\{ \theta \right\} \text{, \ }\Omega \equiv \cup _{\theta
\in \Theta }\Omega ^{\left[ \theta ,\text{ }%
\mathbb{R}
\right] }\text{, }f:\Theta \longrightarrow Z\right] \text{,}
\end{equation*}%
$f$ can be $UD$-implemented by a stochastic infinite-action mechanism, if $%
f\left( \widetilde{\theta }\right) $ is agents' top-ranked outcome for any $%
\widetilde{\theta }\in \Theta ^{\text{una}}\cap \Theta ^{\text{strict}}$.
\end{theo}

\noindent \textbf{Proof of Theorem \ref{thm:UD:infinite}:} For each $i\in 
\mathcal{I}$ , define $\tau ^{i}:\Theta \longrightarrow Z$ such that $\tau
^{i}\left( \widetilde{\theta }\right) $ is the top-ranked outcome for agent $%
i$ at any $\widetilde{\theta }\in \Theta $.

If $f$ is dictatorial with $i$ being a dictator, the $i$-dictatorial
mechanism $UD$-implements $f$. From now on, suppose $f$ is not dictatorial.
We thus have%
\begin{equation}
\theta \notin \left( \Theta ^{\text{una}}\cap \Theta ^{\text{strict}}\right) 
\text{ and }\tau ^{i}\left( \theta \right) \neq f\left( \theta \right) \text{%
, }\forall i\in \mathcal{I}\text{.}  \label{kks1}
\end{equation}%
For each $i\in \mathcal{I}$ , define $\varsigma ^{i}:Z\longrightarrow 2^{Z}$
as:%
\begin{equation*}
\varsigma ^{i}\left( z\right) \equiv \left\{ 
\begin{tabular}{ll}
$\left\{ z\right\} \text{,}$ & if $z\neq f\left( \theta \right) $, \\ 
&  \\ 
$\left\{ z,\tau ^{i}\left( \theta \right) \right\} \text{,}$ & if $z=f\left(
\theta \right) $%
\end{tabular}%
\right. \text{.}
\end{equation*}
$\theta \notin \left( \Theta ^{\text{una}}\cap \Theta ^{\text{strict}%
}\right) $ in (\ref{kks1}) implies%
\begin{equation}
z\neq z^{\prime }\Longrightarrow \left[ \times _{i\in \mathcal{I}}\varsigma
^{i}\left( z\right) \right] \cap \left[ \times _{i\in \mathcal{I}}\varsigma
^{i}\left( z^{\prime }\right) \right] =\varnothing \text{, }\forall \left(
z,z^{\prime }\right) \in Z\times Z\text{.}  \label{kks2}
\end{equation}%
Let UNIF$\left[ Z\right] $ denote the uniform distribution on $Z$. Define $%
\gamma :Z^{\mathcal{I}}\longrightarrow \triangle \left( Z\right) $ as:%
\begin{equation*}
\gamma \left[ \left( z^{i}\right) _{i\in \mathcal{I}}\right] \equiv \left\{ 
\begin{tabular}{ll}
$z\text{,}$ & if $\left( z^{i}\right) _{i\in \mathcal{I}}\in \times _{i\in 
\mathcal{I}}\varsigma ^{i}\left( z\right) $ for some $z\in Z$, \\ 
&  \\ 
UNIF$\left[ Z\right] \text{,}$ & if $\left( z^{i}\right) _{i\in \mathcal{I}%
}\notin \times _{i\in \mathcal{I}}\varsigma ^{i}\left( z\right) $ for any $%
z\in Z\text{,}$%
\end{tabular}%
\right. \text{.}
\end{equation*}%
and $\gamma $ is well-defined due to (\ref{kks2}).

Consider an infinite-action mechanism $M=\left\langle S,\text{ }%
g\right\rangle $ with 
\begin{equation*}
S_{i}\equiv Z\times \mathbb{N}\times Z\text{ and }S\equiv \times _{i\in 
\mathcal{I}}S_{i}\text{,}
\end{equation*}%
and $g\left[ \left( z_{i},n_{i},\widehat{z}_{i}\right) _{i\in \mathcal{I}}%
\right] $ is defined in two cases:

\begin{description}
\item[Case (1):] if $n_{i}=1$ for every $i\in \mathcal{I}$, 
\begin{equation*}
g\left[ \left( z_{i},n_{i},\widehat{z}_{i}\right) _{i\in \mathcal{I}}\right]
=\gamma \left[ \left( z_{i}\right) _{i\in \mathcal{I}}\right] \text{;}
\end{equation*}

\item[Case (2): ] if $n_{i}>1$ for some $i\in \mathcal{I}$,%
\begin{equation}
g\left[ \left( z_{i},n_{i},\widehat{z}_{i}\right) _{i\in \mathcal{I}}\right]
=\sum\limits_{i\in \mathcal{I}}\frac{1}{\left\vert \mathcal{I}\right\vert }%
\times \left( \frac{1}{2n_{i}}\times \gamma \left[ \left( z_{i}\right)
_{i\in \mathcal{I}}\right] +\frac{1}{2n_{i}}\times \text{UNIF}\left[ Z\right]
+\frac{n_{i}-1}{n_{i}}\times \widehat{z}_{i}\right) \text{.}  \label{jaa1}
\end{equation}
\end{description}

We now prove%
\begin{eqnarray}
UD_{i}\left( M,\text{ }u^{\theta }\right) &=&\left\{ \tau ^{i}\left( \theta
\right) \right\} \times \left\{ 1\right\} \times Z\text{, }\forall i\in 
\mathcal{I}\text{, }\forall u^{\theta }\in \Omega _{\theta }\text{,}
\label{taa1} \\
UD_{i}\left( M,\text{ }u^{\widetilde{\theta }}\right) &=&\left[ \varsigma
^{i}\left( \tau ^{i}\left( \widetilde{\theta }\right) \right) \right] \times
\left\{ 1\right\} \times Z\text{, }\forall i\in \mathcal{I}\text{, }\forall 
\widetilde{\theta }\in \Theta ^{\text{una}}\cap \Theta ^{\text{strict}}\text{%
, }\forall u^{\widetilde{\theta }}\in \Omega _{\widetilde{\theta }}\text{,}
\label{taa2}
\end{eqnarray}%
and as a result, $f$ is $UD$-implemented by $M$.

For each $i\in \mathcal{I}$, each $\widetilde{\theta }\in \Theta $, and each 
$u^{\widetilde{\theta }}\in \Omega _{\widetilde{\theta }}$, since%
\begin{equation*}
u_{i}^{\widetilde{\theta }}\left[ \tau ^{i}\left( \widetilde{\theta }\right) %
\right] >\max_{y\in \left[ Z\diagdown \left\{ \tau ^{i}\left( \widetilde{%
\theta }\right) \right\} \right] \cup \left\{ \text{UNIF}\left[ Z\right]
\right\} }U_{i}^{\widetilde{\theta }}\left( y\right) \text{,}
\end{equation*}%
we can find some positive integer $n^{\left( i,u^{\widetilde{\theta }%
}\right) }$ such that%
\begin{equation}
\min_{z\in Z}U_{i}^{\widetilde{\theta }}\left[ \frac{1}{n^{\left( i,u^{%
\widetilde{\theta }}\right) }}\times z+\frac{n^{\left( i,u^{\widetilde{%
\theta }}\right) }-1}{n^{\left( i,u^{\widetilde{\theta }}\right) }}\times
\tau ^{i}\left( \widetilde{\theta }\right) \right] >\max_{y\in \left[
Z\diagdown \left\{ \tau ^{i}\left( \widetilde{\theta }\right) \right\} %
\right] \cup \left\{ \text{UNIF}\left[ Z\right] \right\} }U_{i}^{\widetilde{%
\theta }}\left( y\right) \text{.}  \label{jaa2}
\end{equation}

First, consider state $\theta $. Since $\gamma \left[ \left( z_{j}=\tau
^{i}\left( \theta \right) \right) _{j\in \mathcal{I}}\right] =\tau
^{i}\left( \theta \right) $ achieves the maximal utility for agent $i$ at $%
\theta $, we have 
\begin{equation*}
UD_{i}\left( M,\text{ }u^{\theta }\right) \supseteq \left\{ \tau ^{i}\left(
\theta \right) \right\} \times \left\{ 1\right\} \times Z\text{, }\forall
i\in \mathcal{I}\text{, }\forall u^{\theta }\in \Omega _{\theta }\text{.}
\end{equation*}%
For each $i\in \mathcal{I}$ and each $u^{\theta }\in \Omega _{\theta }$, a
strategy $\left( z_{i}\in Z,n^{i}>1,\widehat{z}_{i}\in Z\right) $ is
strictly dominated by $\left( z_{i},n^{i}+1,\tau ^{i}\left( \theta \right)
\right) $ due to (\ref{jaa1}). Furthermore, a strategy $\left( z_{i}\in
Z\diagdown \left\{ \tau ^{i}\left( \theta \right) \right\} ,n^{i}=1,\widehat{%
z}_{i}\in Z\right) $ is strictly dominated by $\left( z_{i},n^{\left(
i,u^{\theta }\right) },\tau ^{i}\left( \theta \right) \right) $ due to (\ref%
{jaa2}). Therefore,%
\begin{equation*}
UD_{i}\left( M,\text{ }u^{\theta }\right) \subseteq \left\{ \tau ^{i}\left(
\theta \right) \right\} \times \left\{ 1\right\} \times Z\text{, }\forall
i\in \mathcal{I}\text{, }\forall u^{\theta }\in \Omega _{\theta }\text{.}
\end{equation*}%
Therefore, (\ref{taa1})\ holds.

Second, consider any state $\widetilde{\theta }\in \Theta ^{\text{una}}\cap
\Theta ^{\text{strict}}$. Since%
\begin{equation*}
g\left[ \times _{i\in \mathcal{I}}\left( \left[ \varsigma ^{i}\left( \tau
^{i}\left( \widetilde{\theta }\right) \right) \right] \times \left\{
1\right\} \times Z\right) \right] =f\left( \widetilde{\theta }\right) \text{,%
}
\end{equation*}%
which achieves the maximal utility for all agents at $\widetilde{\theta }$,
we have%
\begin{equation*}
UD_{i}\left( M,\text{ }u^{\widetilde{\theta }}\right) \supseteq \left[
\varsigma ^{i}\left( \tau ^{i}\left( \widetilde{\theta }\right) \right) %
\right] \times \left\{ 1\right\} \times Z\text{, }\forall i\in \mathcal{I}%
\text{, }\forall \widetilde{\theta }\in \Theta ^{\text{una}}\cap \Theta ^{%
\text{strict}}\text{, }\forall u^{\widetilde{\theta }}\in \Omega _{%
\widetilde{\theta }}\text{.}
\end{equation*}%
For each $i\in \mathcal{I}$ and each $u^{\widetilde{\theta }}\in \Omega _{%
\widetilde{\theta }}$, a strategy $\left( z_{i}\in Z,n_{i}>1,\widehat{z}%
_{i}\in Z\right) $ is strictly dominated by $\left( z_{i},n_{i}+1,\tau
^{i}\left( \widetilde{\theta }\right) \right) $ due to (\ref{jaa1}).
Furthermore, a strategy $\left( z_{i}\in Z\diagdown \left[ \varsigma
^{i}\left( \tau ^{i}\left( \widetilde{\theta }\right) \right) \right]
,n^{i}=1,\widehat{z}_{i}\in Z\right) $ is strictly dominated by $\left(
z_{i},n^{\left( i,u^{\widetilde{\theta }}\right) },\tau ^{i}\left( 
\widetilde{\theta }\right) \right) $ due to (\ref{jaa2}). Therefore,%
\begin{equation*}
UD_{i}\left( M,\text{ }u^{\widetilde{\theta }}\right) \subseteq \left[
\varsigma ^{i}\left( \tau ^{i}\left( \widetilde{\theta }\right) \right) %
\right] \times \left\{ 1\right\} \times Z\text{, }\forall i\in \mathcal{I}%
\text{, }\forall \widetilde{\theta }\in \Theta ^{\text{una}}\cap \Theta ^{%
\text{strict}}\text{, }\forall u^{\widetilde{\theta }}\in \Omega _{%
\widetilde{\theta }}\text{.}
\end{equation*}%
Therefore, (\ref{taa2})\ holds.$\blacksquare $

\section{Conclusion}

\label{sec:conclude}

This paper resolves a long-standing open question raised by \cite{tb95}: can
randomization, used without money, rescue implementation in undominated
strategies from dictatorship on unanimity-containing domains? The answer is
now precise. If actions are finite and we delete strictly dominated
strategies once, the classic impossibility remains: lotteries by themselves
do not create enough discipline. If we iterate deletion and allow at least
three outcomes, simple transfer-free randomization does create the needed
discipline, and non-dictatorial rules are implementable. If we insist on
one-shot deletion but enrich the action space, infinite actions allow
probability adjustments fine enough to generate strict dominance
outright---again, without quasilinear structure.

Three avenues naturally follow. First, our results show the possibility of $%
UD^{\infty }$-implementation by finite-action mechanisms; a full
characterization of such implementation is of fundamental importance.
Second, clarify the relationship between two notions of rationalizable
implementation---one based on iterative deletion of strictly dominated
strategies and the other based on iterative deletion of never-best
strategies---which coincide on finite-action mechanisms but can diverge with
infinite action spaces; understanding when they agree, when one is strictly
stronger, and how that gap interacts with implementation via stochastic
mechanisms remain open. Third, move to incomplete information: develop a
theory of interim $UD$- and $UD^{\infty }$-implementation, and connect the
analysis to ordinal Bayesian incentive compatibility (see \cite{dmas2004}).
We leave these important questions for future research.

\bibliographystyle{plainnat}

\bibliography{RI}

\end{document}